\documentclass[prb,twocolumn,showpacs,floatfix,epsf,epsfig]{revtex4}
\usepackage{amsmath}
\usepackage{amsfonts}
\usepackage{amssymb}
\usepackage[a4paper, total={6in, 8in}]{geometry}
\usepackage{graphicx}
\usepackage{color}
\DeclareGraphicsExtensions{,pdf,.png,.jpg}

\begin{document}
%work begin on 17.05.2018
%work on 18.05.2018
%work on 23.05.2018
%work on 24.05.2018
%work on 28.05.2018
%work on 30.05.2018
%work on 31.05.2018
%work on 01.06.2018
%work on 04.06.2018
%work on 05.06.2018
%work on 06.06.2018
%work on 07.06.2018
%work on 08.06.2018
%work on 12.06.2018
%work on 13.06.2018
%work on 14.06.2018
%work on 04.01.2019
%work on 05.01.2019
%work on 07.01.2019
%work on 08.01.2019

\title{Prediction of Two-Dimensional Monochalcogenides: MoS and WS}

\author{Dhanshree Pandey$^{1,2}$ and Aparna Chakrabarti$^{1,2}$}

\affiliation {$^{1}$ Homi Bhabha National Institute, Training School Complex, Anushakti Nagar, Mumbai-400094, India}
\affiliation {$^{2}$ Theory and Simulations Laboratory,  Raja Ramanna Centre for Advanced Technology, Indore - 452013, India}

\begin{abstract}
Using density functional theory, we explore  the possibility of two monolayer monochalcogenides, namely, MoS and WS (buckled  and puckered). Our results on cohesive energy and phonon dispersion predict that the buckled structures of both MoS and WS are stable. On the other hand, while the puckered structure of WS clearly shows a dynamical instability, the same for MoS may have a stable configuration. Charge analyses predict ionic-like bonding in these systems. Density of states and band structure reveal a non-magnetic metallic nature for MoS in the stable configurations. However, for the buckled WS, our study predicts a non-magnetic semi-metallic nature. Further, semi-metal to indirect semiconductor transition has been observed for tensile strain of 5\%, 6\% and 8\%.

\end{abstract}
\pacs {71.15.Nc, %Total energy and cohesive energy calculations
~71.15.Mb, %Density functional theory, local density approximation, gradient and other corrections
~68.65.−k} %low dimensional structure
\maketitle
%{\color{blue} }      

\section{Introduction}

For last couple of decades, graphene-like 
two-dimensional ($2D$) monolayer transition 
metal dichalcogenides (TMDCs) are being studied 
extensively by experimentalists and theoreticians 
alike.\cite{PNAS102KSN,Review,SSC12,PhysicaB,PRB79SL,JPCC115CA,ACSNano5SB,PCCP12,NN7QHW,ACSNano8DJ,ACSNano8RG,SciRep6JR,PRB94RKD,JJAP56AO,NM8WW} 
TMDCs of stochiometry MX$_{2}$, where M is a 
transition metal (TM) atom and X is a chalcogen 
atom, exhibit a variety of electronic
properties. These are either insulators or 
semiconductors (TM typically being Ti, Zr, 
Mo, W etc) or metals or semi-metals 
(TM typically being V, Nb etc).
Molybdenum disulfide (MoS$_{2}$) and tungsten disulfide 
(WS$_{2}$) are two of the most typical TMDCs with 
a band-gap and hence are studied in great detail
in the literature.  

%%%literature about bulk and monolayer MoS2 
Bulk structure of MoS$_{2}$ exhibits hexagonal 
symmetry and belongs to the 
space group P6$_{3}$/mmc. 
%Van der Waals interaction (vdW) is known to
%play a significant role in the stabilization of
%these TMDCs.\cite{TMDCvdW}
Since MoS$_{2}$ in bulk form 
is a layered structure,\cite{MoS2bulk} 
a monolayer of MoS$_{2}$ can easily be 
envisaged as a cleaved (001) surface of bulk 
MoS$_{2}$ material.\cite{PNAS102KSN} 
It consists of a buckled
structure and has a three-atom layered 
structure, where one atomic layer of molybdenum 
lies in between two layers of sulfur atoms.

Many studies, interesting from both the points 
of view of practical application  
and fundamental understanding, on monolayer
MoS$_{2}$ are available in the literature.
Non-linear elastic
 behavior of monolayer MoS$_{2}$ system has
been shown by Cooper {\it et al.}\cite{PRB87RCC}
The authors present the in-plane value of 
Young's modulus which is found to match with
the experimental values.\cite{ACSNano5SB} 
They have established
that the system is strong and flexible.
Further, it has been observed that both the 
bulk and its two-dimensional counterpart 
are systems with a finite band-gap and the
latter has a direct band-gap as opposed
to the bulk.\cite{PRB79SL,PRL105KFM,PE56ES}
Monolayer MoS$_{2}$ can be thus seen as 
 a possible candidate for 
flexible electronic devices due to its suitable 
mechanical and electronic properties. Recently, 
it has indeed been used in realizing a low 
power field effect transistor.\cite{NN6BR} 
%The carrier mobility of this system has been
%demonstrated to be lower than that of 
%silicon.\cite{JAP101AA}
To improve or alter the electronic and other
physical properties, studies on effect of strain 
and also vacancy in $2D$ MoS$_{2}$ 
have also been carried out in the 
literature.\cite{NM8WW,NT25KCS,ACSNano5SB,SSC227AEM}

%%%literature about bulk and monolayer WS2
Bulk and monolayer structure 
of WS$_{2}$ are also well-known, like 
 MoS$_{2}$. Extensive theoretical as 
well as experimental studies on this material
as well as other $2D$ TMDCs exist in the 
literature.\cite{WS2etc}

%%%Motivation
%%%why monosulfide?
In this paper, we probe $2D$ monolayer
structures of molybdenum and tungsten chalcogenide. 
As a chalcogen atom, we have probed only sulfur
atom. The other two well-known chalcogen atoms, 
selenium and telurium are known to be somewhat 
toxic in nature. Therefore, 
we have neglected these two atoms, keeping in 
mind the possible inconvenience in 
preparing the samples of 
molybdenum and tungsten monoselenides or 
monotellurides. Our interest in the above mentioned 
materials is due to the following facts.
First of all, numerous interesting and important 
studies have been present in the literature 
in case of both Mo and W disulfides. Therefore, 
we wish to probe the existence of monolayer
monosulfide of Mo and/or W in this study. 
Furthermore, one interesting observation is:
both Mo and W are well-known to have more than 
one oxidation states of 5,4,3,2 in addition to the most 
common valency of 6.
The electronic configuration of sulphur is 
[Ne] 3s$^2$3p$^4$ and thus the valency of sulfur is 
2. When it forms dichalcogenide with
Mo(W), two sulfur atoms amount to total valency 
of 4 and the valency of Mo(W) is considered as 4.
However, in case of monosulfide, the valency of
the anion is only 2 and then the cation 
valency needs to amount to 6. We would like to explore, 
whether this stoichiometry leads to stable 
structures or not, both from energetic and 
dynamic point of view; and if MoS and WS do have 
stable structures, 
what are the geometric configurations these 
materials are likely to possess in the ground 
state, as well as what are the electronic and 
associated properties of the materials.

%%%why buckled and puckered
We probe buckled and 
puckered structures for the monolayer MoS and WS. The buckled structure (having 
two atoms per unit cell)
has been studied since the monolayer 
MoS$_{2}$ and WS$_{2}$ materials 
exhibit buckled structures. On the other
hand, it has been observed in the literature
that some of the $2D$ monochalcogenides 
(namely, CSe, GeSe, SnS etc) possess
the puckered phosphorene-like structures.\cite{IV-VI, IV-monochalcogenide, IV-monochalcogenide1,SnS, GaS-Se, TiBi,GaX, IV-VI1} 
Therefore, in this work, we  
explore both the (buckled and puckered) structures, 
in order to search for the lowest energy
phase of MoS and WS monolayer materials. For the puckered structure, we probe various different positions of Mo(W) with respect to the S atoms \cite{supplementary} and the energetically lowest configuration is reported in this study where Mo(W) occupies the adjacent sites and S occupies the edge sites. We find that the puckered structure has a different geometry compared to the other $2D$ monochalcogenides as reported in Ref.\cite{IV-VI, IV-monochalcogenide, IV-monochalcogenide1,SnS, IV-VI1}

In what follows, we present the methodology 
used in this study, in the next section. 
We discuss our results in the 
following section and finally we summarize and 
conclude in the last section.

\section{Method}  
The equilibrium lattice constants and fractional 
coordinates 
of the systems have been optimized by density 
functional theory (DFT) based electronic structure 
calculations as implemented in Vienna Ab Initio 
Simulation Package 
(VASP)\cite{VASP}. We use the projector 
augmented wave (PAW) method.\cite{PAW}   
For exchange-correlation functional, 
generalized gradient approximation (GGA) given by 
Perdew, Burke, and Ernzerhof
(PBE) over the local density approximation (LDA) 
has been used.\cite{PBE} 
Convergence tests have been performed for energy 
cut-off, $k$ mesh and vacuum distance. We use an 
energy cutoff of 500 eV for the 
planewaves. The final energies have been calculated 
with a $k$ mesh of 31$\times$31$\times$1. We simulate 
the $2D$  buckled and puckered structures of MoS and WS  
using a unit-cell 
geometry with two and four atoms, respectively, 
arranged in a specific 
order (discussed in Results and Discussion section) 
with  a vacuum length
of about 18 \AA, in the z-direction (direction 
perpendicular to the plane of monolayer MoS and WS) 
to avoid the interaction between two
adjacent unit cells in the periodic arrangement.
The energy and the force tolerance for our 
calculations are 1 $\mu$eV and 5 meV/\AA, 
respectively.  
For obtaining the electronic properties of 
the optimized structures, the Brillouin zone 
integration has been carried out using the 
tetrahedron method implemented in 
VASP.\cite{VASP} The binding   
energies ($E_{B}$) have been calculated 
for probing the energetic stability of a material. 
The binding energy is given by, 
$E_{B} = E_{tot} - \Sigma_{i} n_{i}E_{i}^{atom}$, 
where $i$ denotes different types of atoms present 
in the unit cell 
of the material, $E_{tot}$ and $E_{i}^{atom}$ are the 
energy of the system and corresponding atom $i$, 
respectively and $n_{i}$ represents the number of 
same type ($i$) of atoms present in the system. 
These energies have been
analyzed to establish the electronic stability of  
the monolayer systems studied here. 
To calculate the phonon spectra of all the 
materials,  first principles phonon calculations have been performed employing the finite displacement 
method using the VASP\cite{VASP} as well as 
PHONOPY packages.\cite{phonopy}
 A supercell of 6$\times$6 is chosen for the calculation of phonon dispersion. 
 %Convergence is checked by taking 3$\times$3 and 4$\times$4 supercell of monochalcogenides.
 
In order to account for weak dispersion force, if any present in the system, we have also included  the effect of long-range van der Waals 
(vdW) interaction by using the DFT-D2 method \cite{DFT-D2} as implemented in VASP.\cite{VASP} 
The values of $R_0$ (vdW radius) and dispersion coefficient $C_6$ used for the atoms under study  are
listed in Table I.\cite{DFT-D2}

\begin{table}[ht]
\caption{$R_0$ (\AA) and dispersion coefficient, $C_{6} (J nm^{6} mol^{-1})$ for Mo, W and S.}
\label{tab:table1}
\begin{center}
\begin{tabular}{|c|c|c|}
\hline
Atom & $R_0$  & C$_{6}$  \\
\hline
Molybdenum & 1.639 & 24.67 \\
Tungsten   & 1.772 &81.24 \\
Sulfur     &1.683 & 5.57 \\
\hline
\end{tabular}
\end{center}
\end{table}

 %Figure1
\begin{figure*}[h!]
\centering
\includegraphics[scale=0.07]{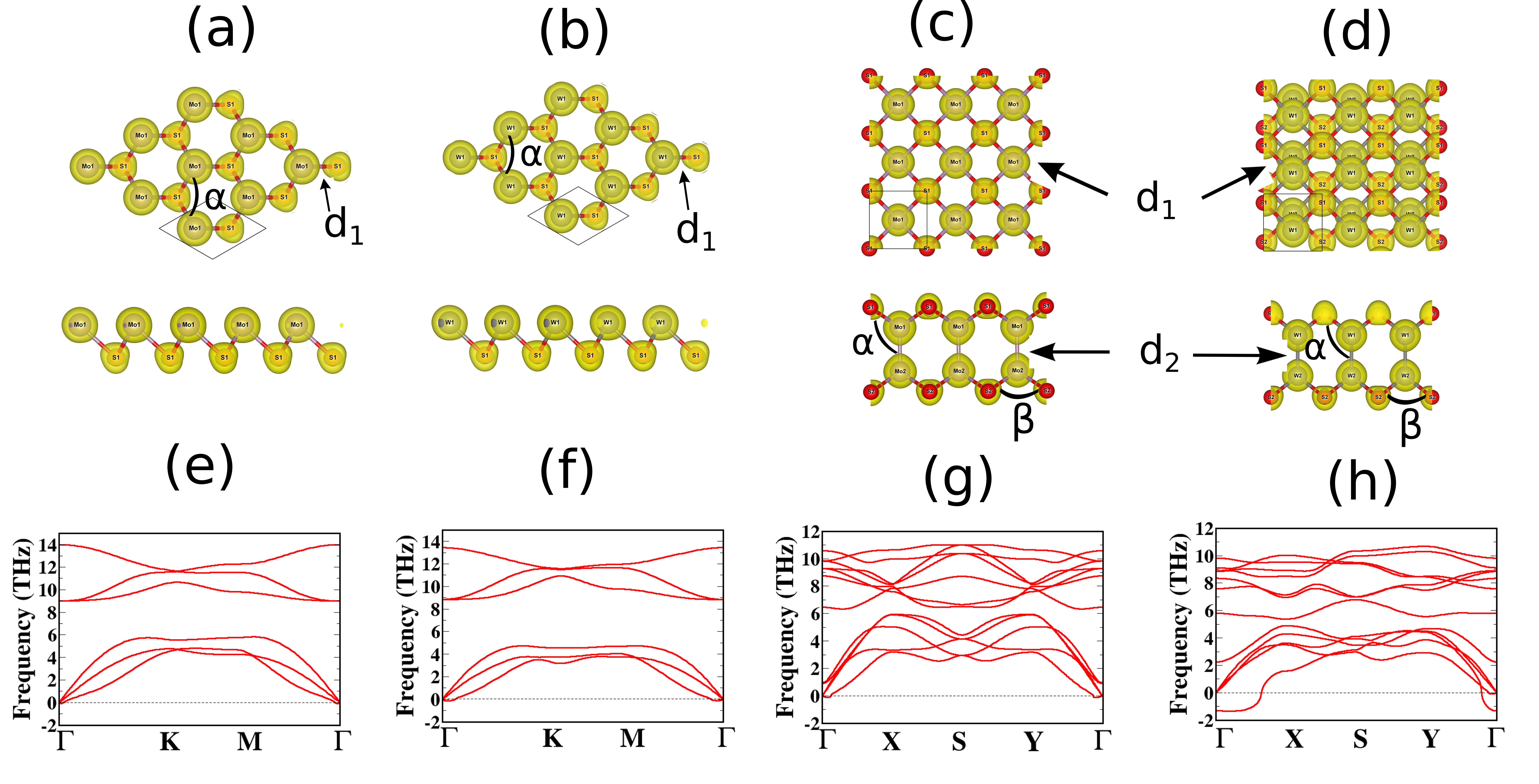}
\caption{Valence charge density distribution (VCDD) for optimized geometry for (a) MoS-buckled, (b) WS-buckled, (c) MoS-puckered and (d) WS-puckered configurations. The solid line indicates the unit-cell. Gray and red balls represent Mo(W) and S atoms, respectively. Yellow region indicates the electron rich region, subsequent to that, the gray color of Mo(W) is not clearly visible. (e), (f), (g) and (h) represent the phonon dispersion curves for MoS-buckled, WS-buckled,  MoS-puckered and  WS-puckered configurations, respectively. WS-puckered case (h) exhibits negative frequency with substantial values, near the $\Gamma$ point.}
\label{fig:Figure1}
\end{figure*}

\section{Results and Discussion}
\subsection{Structure and Energetics}
As mentioned above, we probe two configurations, buckled and 
puckered, for the MoS and WS monolayers. To check and compare the electronic stability of the structures, we calculate the binding energy per atom ($E_B$/atom). From Table II, we note that for both, MoS and WS, puckered configuration is the lowest energy configuration. The difference between the two-configurations for MoS and WS are 0.51 and 0.43 eV/atom,  respectively. Further, valence charge density distribution (VCDD) (Fig.1(a)-1(d))has been studied to predict the nature of bonding and the charge distribution in the system. From VCDD, it is clear that the charges are more localized on atoms which suggests that the nature of bonding is more ionic-like. 
We carried out analysis of Bader charge\cite{Bader} which is expected to provide greater understanding of the bonding nature.
It is to be noted that the Bader charge analysis corroborates with the findings of VCDD. Huge charge transfer is observed
to occur between the X (= Mo, W) and S atoms for all the four monochalcogenides (Table II). The charge transfer is  consistent with the difference in electronegativity of the atoms (as Mo and W has less electronegativity than S). Bader charge analysis also helps in explaining the relative stability of MoS and WS for the same configuration. From  Table II, we note that, more charge transfer occurs in the $2D$ WS system as compared to MoS system. Hence a stronger ionic-type bonding may lead to higher stability of WS systems. It is to noted that the same ionic-like nature is observed for Mo and W dichalcogenides.\cite{PhysicaB} For example, in Mo$S_2$, 0.90e is lost by Mo and each S gains 0.45e, \cite{PhysicaB}, these values are close to those observed by us in case of MoS.
To gain more insight into the nature of bonding, we carry out analysis of electron localization function (ELF), presented in the supplementary material\cite{supplementary} which further supports the results obtained from VCDD and Bader charge analysis.

It is well known that vdW interaction, in spite of its weak strength,  plays an important role in stabilizing several multi-layered systems\cite{vdw-layered-sci-paper, vdw1, vdw2, vdw3}. Probing the vdW interaction in the monolayers studied here can give us an insight about the interactions present in the systems. Therefore, vdW dispersion corrected calculations have been carried out for all the systems under study. Table II lists all the geometrical parameters (in the parantheses) obtained from the vdW corrected calculations as well. We observe that the lattice constants reduce in WS cases as a result of vdW forces acting on the atoms, indicating presence of a slightly stronger binding among the atoms in these systems. Further, it is noted that the energies of all the systems get slightly lowered (by a sub-eV range). This signifies that the vdW  may only have a small contribution to the overall interaction present in these systems. The trend of the ordering of the configurations, in terms of energetics, however, remains unaltered. The rigidity and hence structural stability is also checked by analyzing the phonon-dispersion for all the systems under study. A brief discussion on the  phonon modes for these systems will be discussed in the following section.

\begin{table*}[ht]
\caption{Geometric and energetic data for different configurations of MoS and WS monochalcogenide. $E_B$/atom and  $a$ represents the binding energy per atom (in eV/atom) and the lattice constants respectively. $\Delta$ is the buckling parameter, $d_{1}$ and $d_{2}$ represent X-S and X-X bond distances, respectively. Values of $a$, $\Delta$,  $d_{1}$ and $d_{2}$ are in Angstrom unit. $\alpha$ and $\beta$ represent the bond angles S-X-X and S-X-S, respectively (in degrees), where X = Mo, W. $\Delta Q$ shows the amount of Bader charge transfer between X and S atoms in the units of e. Values in parentheses are results from calculations with vdW interaction. Charge analysis is carried out only without vdW correction.}
\label{tab:table2}
\begin{center}
\begin{tabular}{|c|c|c|c|c|c|c|}
\hline
Structure & $E_B$/atom  & $a$  & $\Delta$ & $d_1$, $d_2$  & $\alpha$, $\beta$  & $\Delta Q  $  \\
\hline
MoS-buckled & -4.948949 & 3.000 & 1.577 & 2.342 & 79.6 & 0.50 \\
& (-5.125175)& (3.000) & (1.577)& (2.342) & (79.6) &\\
WS-buckled & -5.838209 & 2.976 & 1.617 & 2.658 & 78.18 & 0.55 \\
& (-6.123231)& (2.962) & (1.618) & (2.354) & (77.9)& \\
MoS-puckered & -5.461969 &  3.121 & - & 2.455, 2.356 & 116.1, 78.9 & 0.76\\
& (-5.729954) & (3.121) & (-) &(2.449, 2.325) & (115.6, 79.1) &\\
WS-puckered & -6.266106 & 3.08, 3.01 & - & 2.448, 2.472 & 105.9, 77.9 & 0.78\\
& (-6.807773) & (3.014, 2.974)& (-)& (2.431, 2.443)& (107.2, 76.6)& \\
\hline
\end{tabular}
\end{center}
\end{table*}

\subsection{Phonon Dispersion}
At T = 0K, a crystal lattice is in its ground state. Hence there is no contribution of phonons in the system. However, at a finite temperature, study of phonons is important as they play crucial role in governing the thermal and electrical properties of a material.
Moreover, the stability of a structure is also dependent on its phonons modes.  
For a stable structure, all the phonon frequencies should be positive.\cite{phonon1} Here, we show the phonon dispersion along the high symmetry path (Fig.1(e)-1(h)) for all the four structures studied in this paper and comment on the structural stability of the systems.

{\it Buckled cases}: Since we know that there exist a total of 3N number of phonon modes, where N is the number of atoms in the unit cell of a system, six phonon modes are present in case of MoS and WS buckled configurations (as the number of atoms is two in these two cases). The lower three modes correspond to the acoustic and the other three to the optical modes. All the modes are having positive frequencies except one mode which has negligibly small value of imaginary frequency at the $\Gamma$- point. This very small negative frequency may be considered as an effect of computational parameters as already observed in the literature.\cite{phonon, arsenene} For MoS-buckled case, there are three optical modes, in which two have values 9 THz (doubly degenerate) and one has a value of 14THz at $\Gamma$-point.  On the other hand, for WS-buckled case, the three optical modes are shifted slightly below as compared to the case of MoS.

{\it Puckered cases}: The puckered configuration for MoS and WS contain a total of 12 phonon modes. From  Fig.1(h), we   see that one of the phonon modes of WS has reasonably large imaginary frequency and thus this system is likely to lack the structural stability. The imaginary frequency can be attributed to the lower symmetry structure for puckered WS configuration.  However, puckered MoS is found to have a stable structure as all the phonon modes are positive.
Since puckered configuration for WS does not show structural stability, this system will not be considered for further discussion. 

Regarding the possibility of growing 2D mono or dichalcogenide materials, we note that many techniques are available in the literature for fabricating these materials.\cite{review1} For example, fabrication of monolayer PtSe$_2$ by direct selenization of Pt has been reported by Wang et al.\cite{PtSe2} Further, monolayer CuSe has also been grown using similar technique.\cite{CuSe} Therefore, sulphidization on Mo and W metal substrates could be a probable method to grow the MoS and WS structures predicted in this study.

%Figure2
\begin{figure*}[h!]
\centering
\includegraphics[scale=0.15]{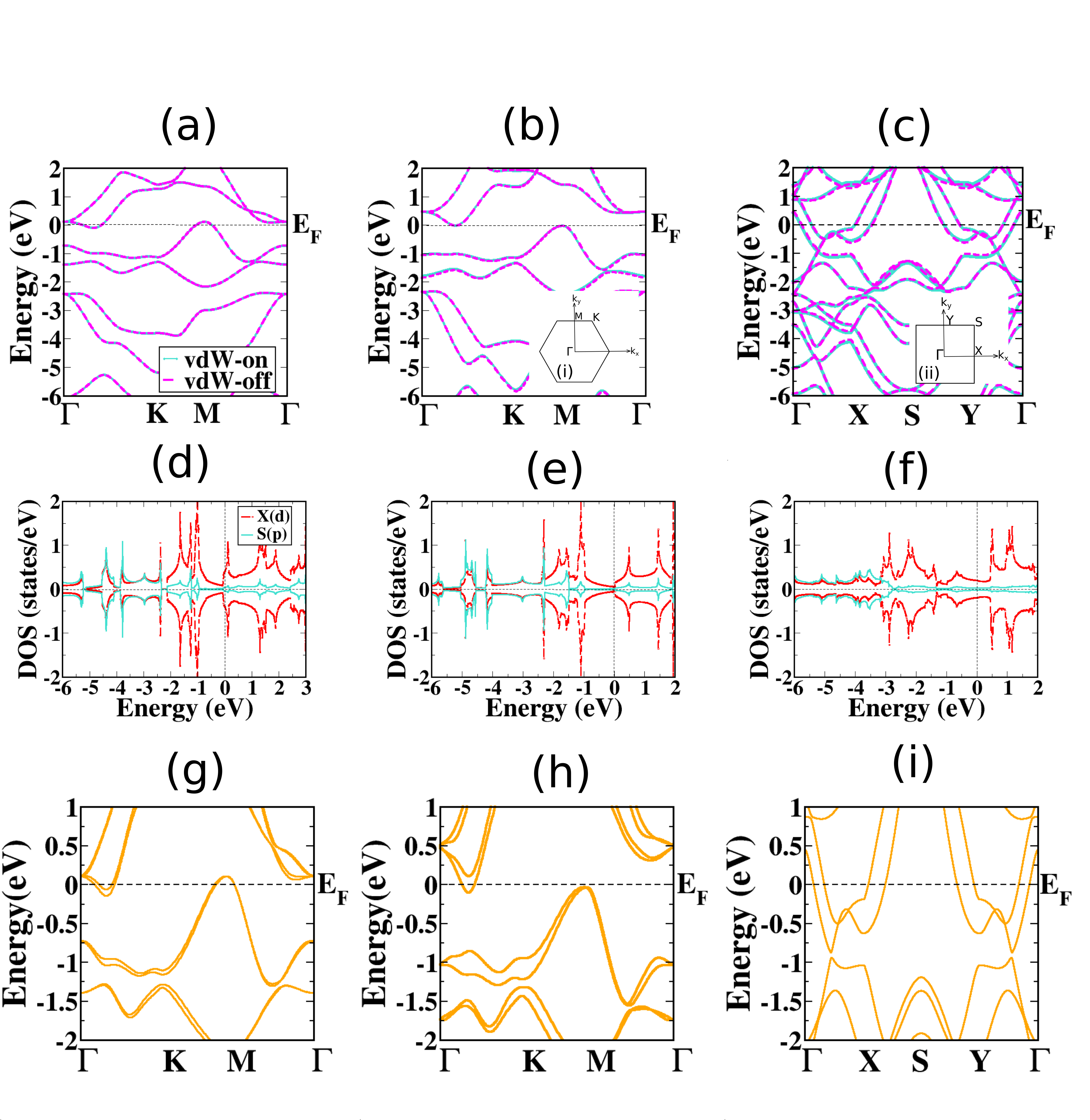}
\caption{Figure presents the results of band structure, partial density of states (PDOS) and bands with spin-orbit interaction. (a)-(c) shows the band structure plot with and without van-der Waals correction, (d)-(f) represent the PDOS and (g)-(i) exhibit the band with the inclusion of spin-orbit coupling. X in the figure reperesents Mo and W. MoS-buckled, WS-buckled and MoS-puckered cases are shown in left, middle and right panels respectively. (i) and (ii) in the inset of (b) and (c), respectively, show the Brillouin zone for the buckled and the puckered system, respectively.}
\label{fig:Figure2}
\end{figure*}

\subsection{Electronic Structure}
In this section, we present the results of electronic band structure and density of states (DOS) (Fig.2) for the monochalcogenide systems. We also analyze the partial DOS (PDOS) to understand the role of orbitals of different atoms in the system. We first discuss the results of the buckled-monochalcogenides followed by the puckered-cases. 

\textbf{Buckled MoS} is found to be a non-magnetic metallic system. It is apparent from the PDOS (Fig.2(a), 3(a)) that the Mo-d states has major contribution around Fermi energy ($E_F$) and hence the band edges comprise of Mo-d derived states with small contribution from S-p orbitals. However, below -2eV, Mo-d and S-p states are strongly hybridized. Fig.3 shows the orbital projected DOS with the partial charge density (PCD) (for C$\Gamma$-point, i.e. the conduction band at the $\Gamma$-point) for MoS. It is clear from Fig.3 that the bands in the region closer to the Fermi level has major contribution from $d_{{x^2} - {y^2}}$ and $d_{xz}$ of Mo atom  with very small contribution of S-p orbitals. PCD reveals the same nature of orbitals contributing to the DOS and bands slightly above the Fermi level. 

In the literature it is found that monolayer MoS$_{2}$ and WS$_{2}$ has a direct band-gap, in contrast to the corresponding  monochalcogenide case.\cite{PhysicaB} Therefore, we compare the two band structures, one of XS$_{2}$ and XS where X is Mo and W as shown in Fig.4. It is to noted that the band gap obtained for MoS$_{2}$ and WS$_{2}$ are 1.65 eV and 1.84 eV respectively, which is consistent with the literature\cite{PhysicaB}. From the figure, we observe a band-gap in MoS$_2$ which disappears in MoS. Moreover, the valence band maxima (VBM) and the conduction band minima (CBM) are at the same k-point for MoS$_2$ which is not so in the MoS case. Further, we observe Mo-d bands in the conduction region closer to the Fermi level for MoS case compared to the corresponding  dichalcogenide case. The clear difference in the electronic structure between the mono and dichalcogenides can be attributed to the absence of one S anion and hence the difference in hybridization between the cation and the anion in the MoS monochalcogenide system. 

For the \textbf{ buckled WS} case, the electronic structure calculation predicts a non-magnetic indirect-type semi-metallic nature for WS-buckled configuration as the VBM (which lies at the $M$-point) and CBM lie at different k-points. Similar to the MoS-buckled case, the contribution to the DOS around $E_F$ is dominated by the W-d states and there is a stronger hybridization between W-d and S-p states farther below $E_F$. However, unlike the corresponding MoS case, the d-states just above the Fermi-level have contribution from also the S-p orbitals along with the W-d states. This is justified by the respective PCD which in this case exhibits charge density around the S-atom as well. Similar to the comparative analysis of MoS$_2$ and MoS, we observe a rearrangement of the bands in WS monochalcogenide case also. This leads to a change in the electronic nature due to the difference in W-S hybridization between the mono and dichalcogenides.

Effect of {\bf spin-orbit coupling (SOC)} has also been probed as it may play an important role in modifying the electronic properties of the system as observed in the literature.\cite{SOC1,SOC11,SOC2, SOC3, SOC4} Wang \textit {et al.} \cite {SOC3} have reported that WS$_2$ used as a substrate opens a path for topological states of matter in graphene-based systems. It is well-known that SOC lifts the degeneracy of energy levels and hence causes the splitting of the levels. Fig.2(g)-2(i) shows the splitting of the electronic levels for the monochalcogenides. 
Furthermore, the inclusion of SOC changes the electronic structure of WS ( Fig.2(h)) such that we observe a slight splitting of the band corresponding to CBM along the $\Gamma$-K path. The more appreciable effect of SOC in WS-buckled compared to the MoS case is expected as the spin-orbit interaction is more pronounced in the elements having higher atomic number.

For \textbf{puckered MoS}, the band structure and the DOS calculations reveal a non-magnetic metallic nature of the system with Mo-d electrons contributing to the states near $E_F$. Effect of SOC on electronic structure has been probed and very small splitting is observed along $\Gamma$-X and $\Gamma$-Y path (as they are symmetric) and also at the S-point (at around -1.2 eV).

The band structures with {\bf van der Waals interaction} are also plotted in Fig.2((a)-(c)), (dashed blue lines). It is observed that the vdW corrected band structures, are very similar in terms of the band dispersion and band energies when compared with the bands without vdW correction. Overall, it can be inferred that the vdW interaction is not likely to play a crucial role in these systems.

%Figure3
\begin{figure}[h!]
\centering
\includegraphics[scale=0.20]{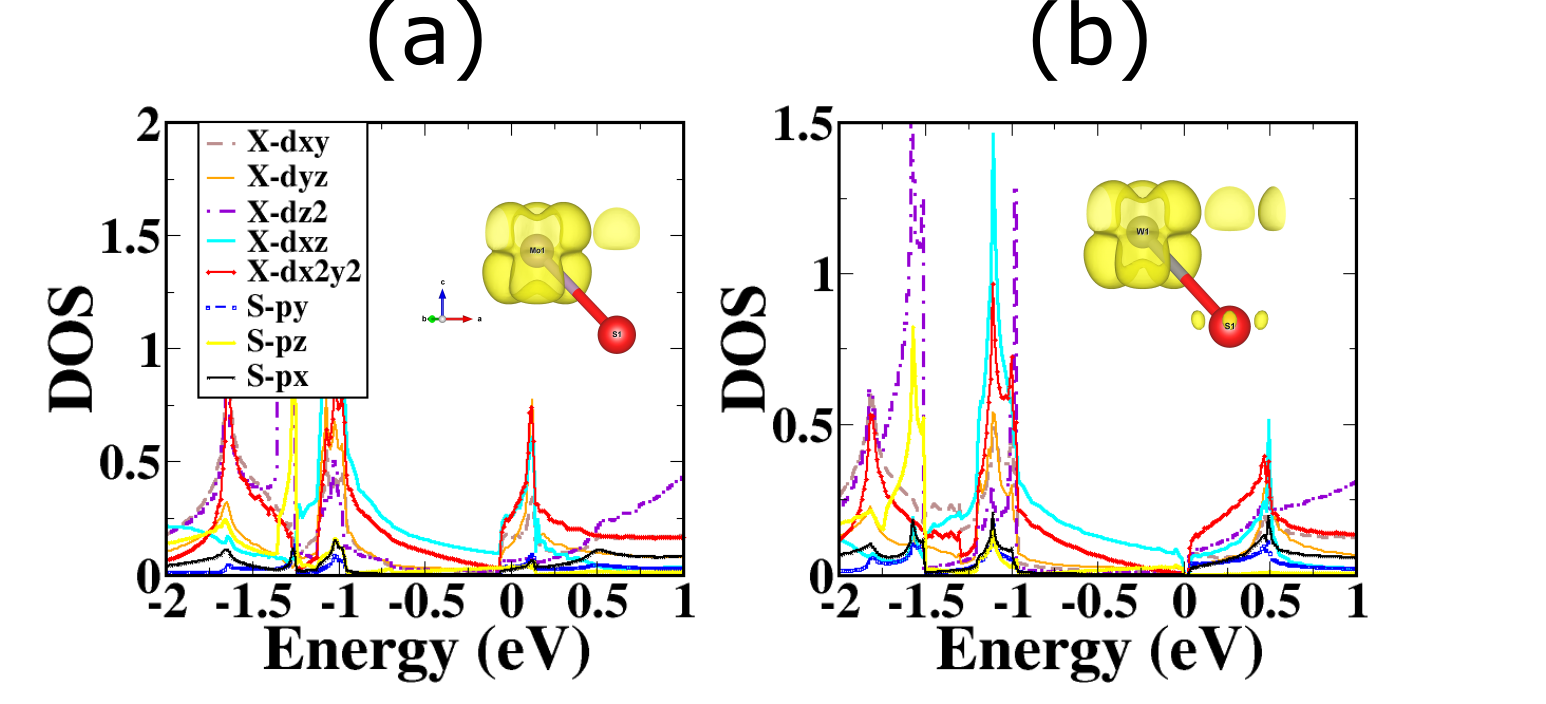}
\caption{Orbital projected density of states (DOS) for buckled monolayer of (a) MoS and (b) WS. DOS is in units of states/eV. Inset shows partial charge densities corresponding to band point C$\Gamma$ located in Fig.4(b,d). }
\label{fig:Figure3}
\end{figure}

%Figure4
\begin{figure}[h!]
\centering
\includegraphics[scale=0.22]{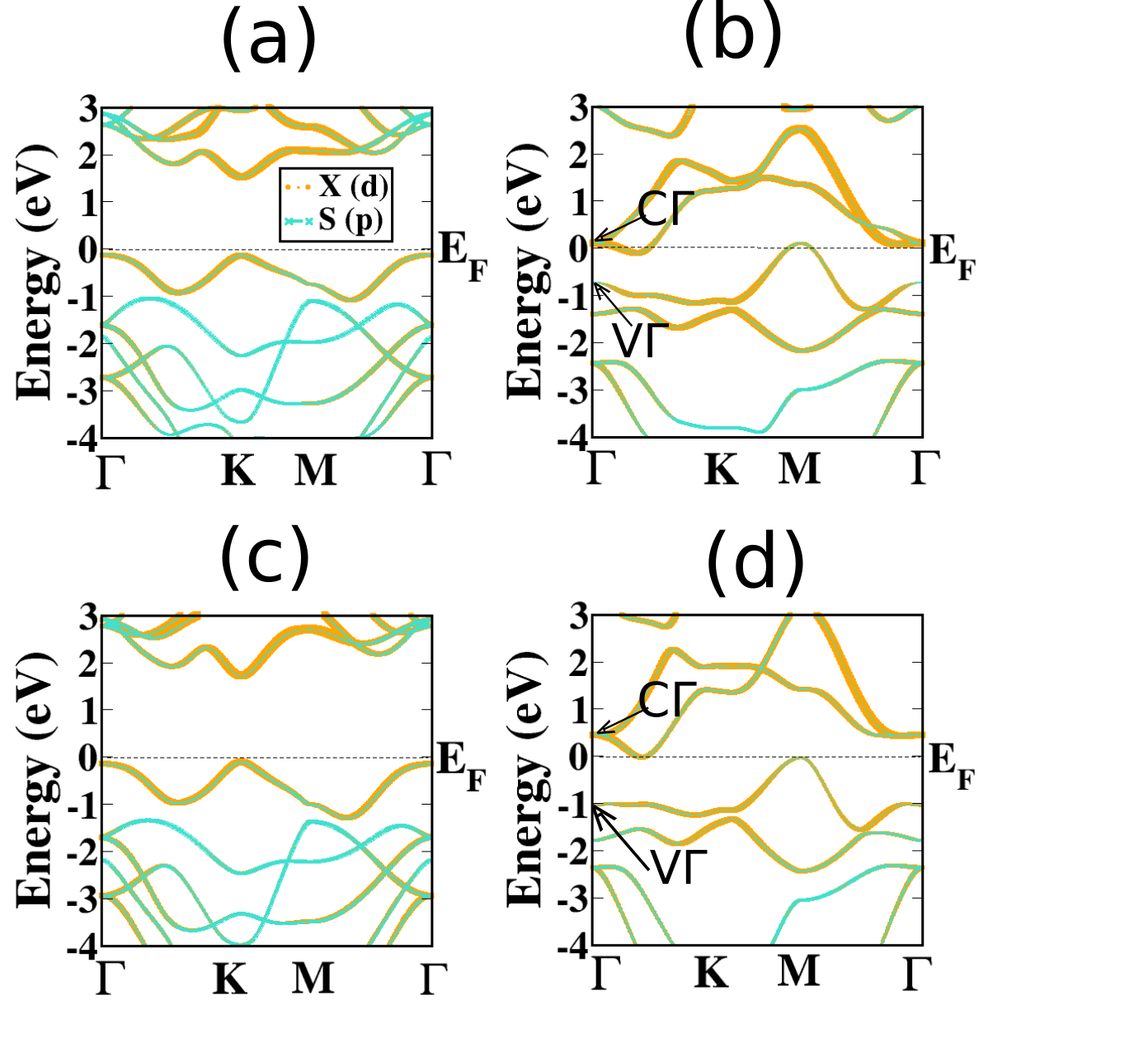}
\caption{Atom projected partial band structure for buckled monolayer of (a) MoS$_2$ and (b) MoS, (c) WS$_2$ and (d) WS. X in the legend corresponds to Mo and W atoms. C$\Gamma$ and V$\Gamma$ represents the $\Gamma$-point in the conduction and valence band regimes, respectively.}
\label{fig:Figure4}
\end{figure}

\subsection{Band-gap engineering by application of strain}
From the literature, it is observed that application of strain (tensile or compressive) plays a major role in the engineering of band-gaps in the material.\cite{PCCP12, ACSNano5SB,NM8WW,SSC227AEM, strain1, strain2, strain3}
Since WS buckled structure is found to be an indirect-type semi-metal, it may be interesting to probe its  electronic structure after application of strain. 
The strain $\delta$  can be defined as $\delta$ = ((a-a$_0$)/a$_0$) $\times$ 100\%, 
where a and a$_0$ are the lattice constants of strained and unstrained systems, respectively. For each value of bi-axial strain, all atoms are  relaxed with the fixed lattice constant. We study the band evolution for both tensile as well as compressive bi-axial strain, with the amount of strain applied to be 5\%, 6\%, and 8\%.
 In the bi-axial strain, we deform the in-plane lattice parameters $a$ and $b$ by an equal amount. The deformation obviously leads to changes in W-S distances, bond angles and the buckling parameter and hence in the electronic structure. From Fig.5, we observe that for the tensile bi-axial strain, a band-gap opens up. However, the band-gap obtained is found to be indirect. Occurrence of a gap is not observed for the compressive strain. 
The origin of the gap can be explained by the evolution of VBM (at $M$-point) and CBM (along $\Gamma$-K path). We observe that with the 
increase in the bi-axial tensile strain, the VBM at $M$-point shifts below the $E_F$ level and the CBM (along $\Gamma$-K path) shifts away from  $E_F$.  The observed shifting of bands has the following origin. From Fig.2(b)and 2(e), it is apparent that the band edges comprise of predominantly the W-d  states with a slight contribution from S-p states. The increase in tensile strain causes the lattice constant (therefore the W-W, W-S or S-S distances) (Table III) to increase. Consequently, the  hybridization of W-d states and S-p states is decreased (Fig.7(b)) which leads to a gap. It is to be noted that C$\Gamma$ (0.46eV above the Fermi level) and V$\Gamma$ (at -1 eV) points are lowered and raised, respectively, towards the Fermi level with the increase in the tensile strain. However, +8\% tensile strain is not sufficient to cause the Fermi level crossing of bands corresponding to $\Gamma$-point and thus we observe an indirect semiconductor behavior of buckled WS monochalcogenide. On the other hand, if we observe the band evolution for the compressive strain, the reverse is observed. CBM and VBM crosses the Fermi level leading to a metallic state.

It is to be noted that unlike the WS case, for buckled MoS, application of strain does not show any metal-semiconductor transition. For the sake of comparison with the WS case, we discuss here the buckled MoS under tensile strain (Fig.6). Similar to the case of WS, the hybridization between Mo-d and S-p states is reduced (Fig.7(a)). The VBM is shifted below $E_F$ and the CBM shifts away from the Fermi level upward when tensile strain is applied to buckled MoS. However, along with this, the  C$\Gamma$ (0.13 eV above the Fermi level) and V$\Gamma$ (at around -0.73 eV) (Fig.4) points also shift  towards the Fermi level. In resemblance with the case of WS,  C$\Gamma$ and V$\Gamma$ points move closer to $E_F$ on the  application of a tensile strain. Since these two points are closer to $E_F$ as compared to that in WS case, it is apparent from Fig.7 that the shift of the bands at $\Gamma$-point (i.e. C$\Gamma$ and V$\Gamma$) towards the Fermi level is responsible for retaining the metallic behavior in this case.

%We analyze the evolution of band for  C$\Gamma$-point. For the unstrained system, C$\Gamma$-point lies around 0.13eV above the Fermi level. With the increase of tensile strain, the C$\Gamma$-point moves downward closer to $E_F$. Moreover, the band corresponding to VBM at $\Gamma$-point also shifts towards Fermi level causing the crossing of the Fermi level. It is interesting to note that VBM (at $M$-point) and CBM (along $\Gamma$-K path) shifts to $\Gamma$-point on application of tensile strain.  From Fig.7, it is apparent that at +8\% strain, the movement of the bands at $\Gamma$-point closer to Fermi level is responsible for retaining the metallic behavior of the buckled MoS. This shifting of bands at $\Gamma$-point also occurs for WS case. However, the C$\Gamma$-point in WS case is around 0.46eV above the Fermi level. Thus, even though there is a shifting of  bands corresponding to $\Gamma$-point closer to $E_F$, +8\% tensile strain is not sufficient to cause the Fermi level crossing of bands corresponding to $\Gamma$-point and thus we observe an indirect semiconductor behavior of buckled WS monochalcogenide.    

It is worth-mentioning here that we have also calculated the binding energy per atom for the strained  MoS and WS systems (Fig.7(c)). We find a parabolic behavior of energy as a function of the bi-axial strain. The unstrained system possesses the lowest energy. However, it is clear from  Fig.5 that overall the tensile strained systems have comparatively higher binding energy (more negative) as compared to that of compressive strained systems. Thus from the energetics, it can be argued that the ease to stretch the system is more than to compress it. 

Like silicon and germanium which are indirect band-gap semiconductors and used in the field of semiconductors, strain-applied indirect band-gap semiconductor buckled WS may be expected to find potential application in the semiconductor industry.
However, though we arrive at an  indirect semiconductor on application of tensile strain, it is to be noted that, the calculations have been performed at T = 0K. Effect of ambient conditions and the substrate effect can alter the electronic structure which requires further investigation and is beyond the scope of the present work.

\begin{table}[ht]
\caption{Lattice constant (a), buckling parameter ($\Delta$), W-S bond distance ($d_1$), W-S-W bond angle ($\alpha$) and the band-gap ($E_g  $) are tabulated for different tensile bi-axial strain. a, $\Delta$, $d_1$ are in \AA, $\alpha$ in deg, $E_g  $ is in eV units. }
\label{tab:table2}
\begin{center}
\begin{tabular}{|c|c|c|c|c|c|c|}
\hline
Strain (\%) &  a   &  $\Delta$  & $d_1$  & $\alpha$   & $E_g  $   \\

\hline
5\% &  3.124 &  1.556 & 2.382 & 81.9 & 0.14 \\
6\% & 3.154 & 1.545 & 2.388 & 82.7 & 0.17\\
8\% & 3.213 & 1.520 & 2.40 & 84.1 & 0.20\\

\hline
\end{tabular}
\end{center}
\end{table}

%Figure5
\begin{figure}[h!]
\centering
\includegraphics[scale=0.135]{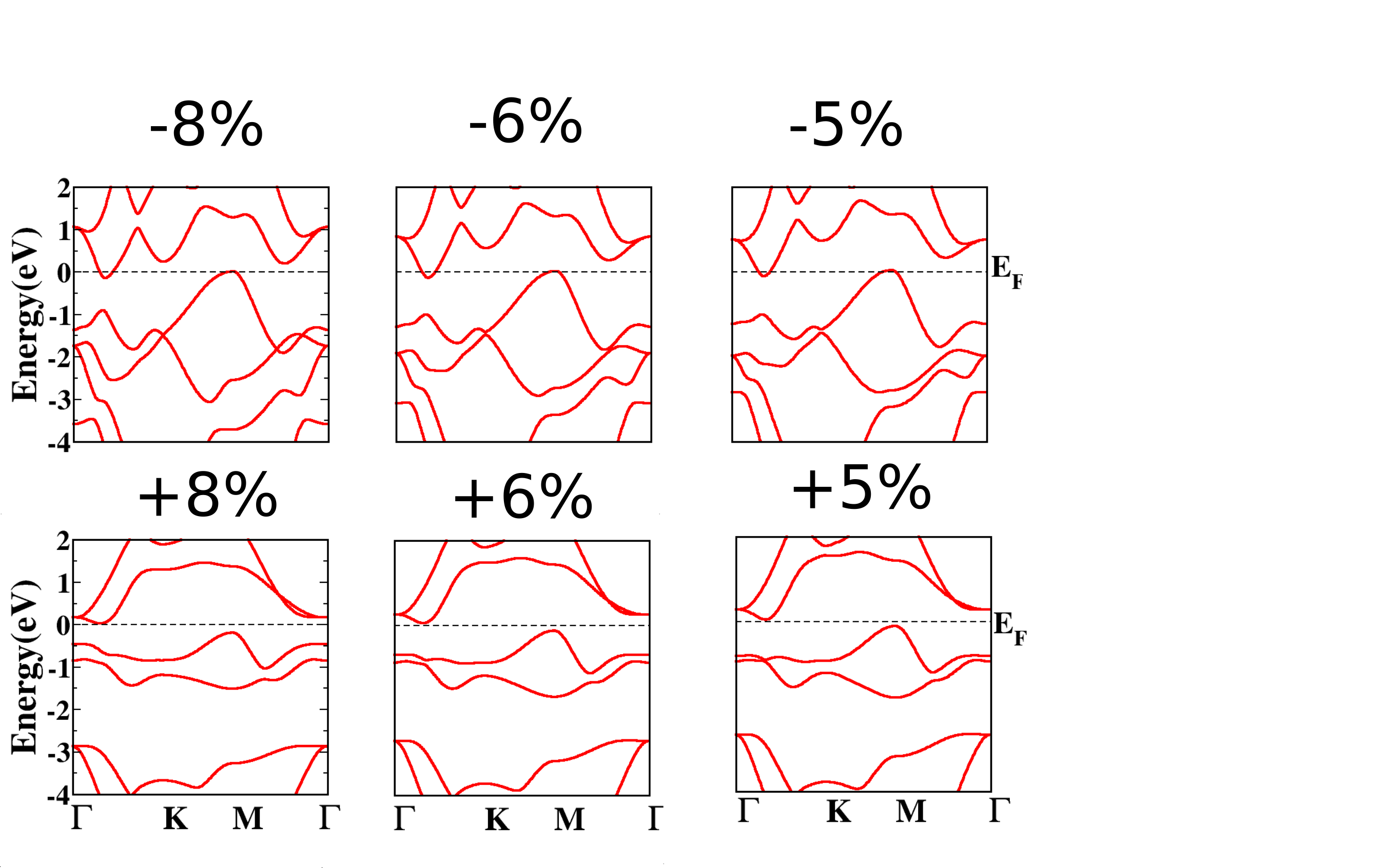}
\caption{Evolution of band-structure due to bi-axial tensile and compressive strain on WS-buckled system.Tensile strain causes the opening of band-gap.}
\label{fig:Figure5}
\end{figure}

%Figure6
\begin{figure*}[h!]
\centering
\includegraphics[scale=0.18]{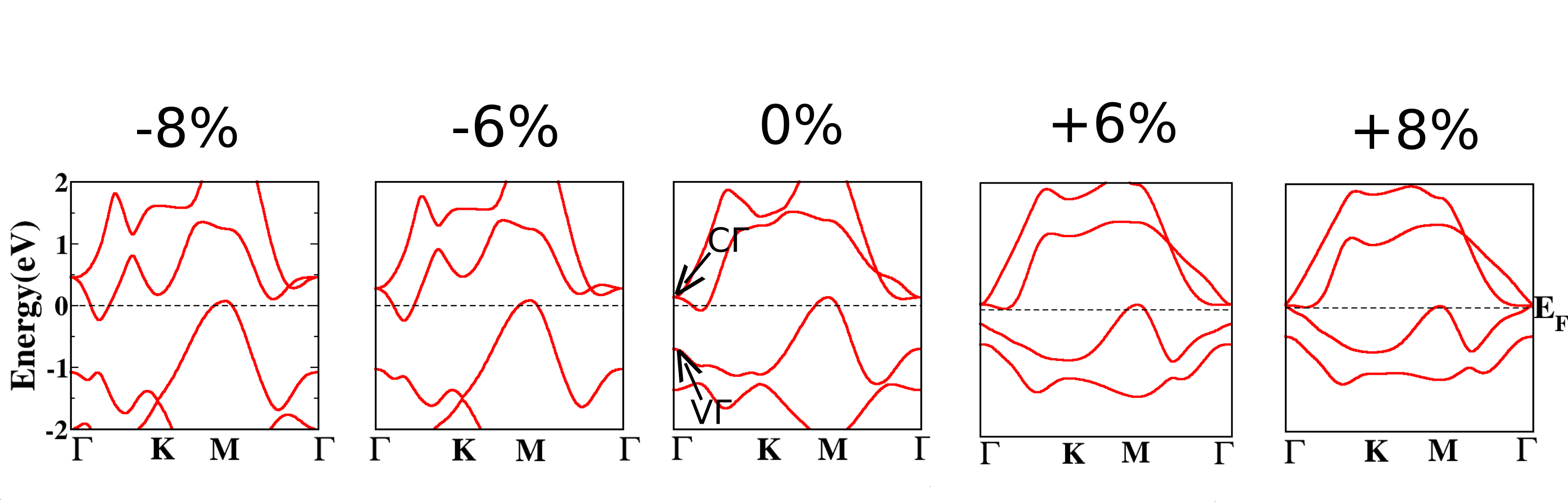}
\caption{Evolution of band-structure due to bi-axial tensile and compressive strain on MoS-buckled system.}
\label{fig:Figure6}
\end{figure*}

%Figure7
\begin{figure*}[h!]
\centering
\includegraphics[scale=0.25]{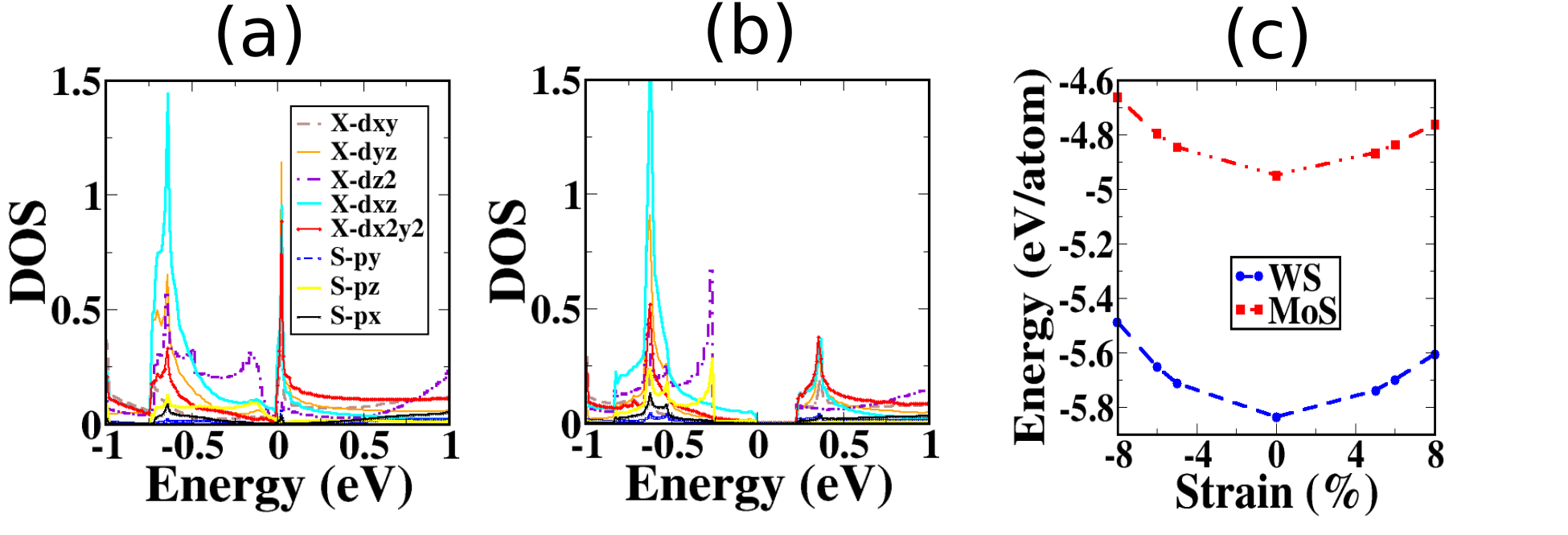}
\caption{Orbital projected DOS (in units of states/eV) corresponding to +8\% tensile strain for (a) MoS and (b) WS. (c)Binding energy per atom (eV/atom) of  monolayer buckled MoS and WS as a function of strain.}
\label{fig:Figure7}
\end{figure*}

\section{Conclusion}
Using density functional theory based {\it ab initio}
calculations (VASP programme package), we predict 
two new monolayered monochalcogenides, namely,
 molybdenum monosulfide (MoS) and tungsten 
monosulfide (WS). We carry 
out full geometry optimization of two different
configurations for the MoS and WS monolayer 
structures, buckled and puckered. From the structural and energetic stability point of view, we found that MoS is stable in both the configurations, the puckered being more stable. However, WS in puckered configuration, though energetically the most stable one, is found to be dynamically unstable. Further, effect of vdW has been studied which suggest negligible change in the electronic properties of the stable structures. VCDD and Bader charge analysis predicts an ionic-like bonding in the system. DOS and electronic band structure shows the non-magnetic metallic nature for MoS in both the stable configurations. On the other hand, for the buckled WS, our study reveals the non-magnetic indirect semi-metallic nature. The difference in the electronic nature of MoS and WS from their dichalcogenide analogue i.e. MoS$_2$  and WS$_2$ (which are both direct gap semiconductor) can be attributed to the absence of one sulphur atom and hence the difference in Mo(W)-S hybridization. 
Effect of spin-orbit coupling (SOC) has also been probed. It is observed that, since W has higher atomic number, the effect of SOC is slightly more pronounced in the case of WS, compared to MoS. 

Finally, the effect of bi-axial tensile and compressive strain on the electronic structure has also been explored. Interestingly, a transition from semi-metal to an indirect band-gap semiconductor has been observed for the tensile strain of 5\%, 6\% and 8\% in case of buckled WS. Thus, it is predicted that, with a suitable choice of a substrate, WS system may turn out to be important in the field of semiconductors.

\section{Acknowledgement}
Authors thank P. A. Naik  for facilities and encouragement. D.P. thanks C. Kamal for fruitful discussions. The scientific 
 computing group, computer centre of RRCAT, Indore  is thanked for the help in installing of and support in running the codes. D.P. thanks HBNI and RRCAT for financial support.

{}

\end{document}